\documentclass[%
 aip,amsmath,amssymb,
 reprint,
]{revtex4-1}

\usepackage{graphicx}
\usepackage{amsmath}
\usepackage{dcolumn}
\usepackage{bm}

\usepackage[utf8]{inputenc}
\usepackage[T1]{fontenc}
\usepackage{newtxtext}
\usepackage{newtxmath}
\usepackage{etoolbox}

\usepackage{graphicx,dcolumn,bm,enumerate,eurosym}
\usepackage{amsfonts,amsmath,amsopn,amssymb}
\usepackage{graphicx,graphics}
\usepackage{color,verbatim,mathrsfs,braket,xargs,lipsum}
\usepackage[colorlinks=true, allcolors=blue]{hyperref}
\usepackage{mathtools}
\usepackage[dvipsnames]{xcolor}

\usepackage[normalem]{ulem}

\hypersetup{
    colorlinks,
    linkcolor={blue!50!blue},
    citecolor={blue!50!blue},
    urlcolor={blue!80!black}
}

\usepackage[normalem]{ulem} 
\definecolor{mgreen}{RGB}{1,123,0}

\usepackage{graphicx,dcolumn,bm,enumerate,eurosym}
\usepackage{amsfonts,amsmath,amsopn,amssymb}
\usepackage{graphicx,graphics}
\usepackage{color,verbatim,mathrsfs,braket,xargs,lipsum}
\usepackage[colorlinks=true, allcolors=blue]{hyperref}
\usepackage{mathtools}
\usepackage[dvipsnames]{xcolor}

\usepackage[normalem]{ulem}

\hypersetup{
    colorlinks,
    linkcolor={blue!50!blue},
    citecolor={blue!50!blue},
    urlcolor={blue!80!black}
}

\usepackage[normalem]{ulem} 
\definecolor{mgreen}{RGB}{1,123,0}

\makeatletter
\def\@email#1#2{%
 \endgroup
 \patchcmd{\titleblock@produce}
  {\frontmatter@RRAPformat}
  {\frontmatter@RRAPformat{\produce@RRAP{*#1\href{mailto:#2}{#2}}}\frontmatter@RRAPformat}
  {}{}
}%
\makeatother
\begin{document}

\preprint{AIP/123-QED}

\title[Non-Markovian quantum-emitter dynamics and bound-state formation in an Aubry--Andr\'{e}--Harper photonic lattice]{Non-Markovian quantum-emitter dynamics and bound-state formation in an Aubry--Andr\'{e}--Harper photonic lattice}

\author{Zahra Jalali-Mola}
\affiliation{School of Physics and CRANN Institute, Trinity College Dublin, The University of Dublin, Dublin 2, Ireland}
\email{jalali.mola@gmail.com}

\author{Jule Neidhart}
\affiliation{School of Physics and CRANN Institute, Trinity College Dublin, The University of Dublin, Dublin 2, Ireland}

\author{Saeid Asgarnezhad-Zorgabad}
\affiliation{School of Physics and CRANN Institute, Trinity College Dublin, The University of Dublin, Dublin 2, Ireland}

\author{Ortwin Hess}
\affiliation{School of Physics and CRANN Institute, Trinity College Dublin, The University of Dublin, Dublin 2, Ireland}
\affiliation{Rinn Quantum, School of Physics, Trinity College Dublin, The University of Dublin, Dublin 2, Ireland}
\email{ortwin.hess@tcd.ie}

\date{\today}

\begin{abstract}
In non-Markovian light--matter interactions, a spectrally structured bath can alter the evolution of a quantum emitter by modifying the vacuum fluctuations experienced by the emitter.
This raises the question of how the localization transition and the associated multifractal critical modes induced by a quasiperiodic potential shape the emitter dynamics.
We investigate this scenario in the single-excitation sector for a two-level quantum emitter locally coupled to an Aubry--Andr\'e--Harper (AAH) photonic lattice.
By externally tuning the strength of the quasiperiodic onsite modulation, the lattice is tuned from the extended to the localized regime through a critical point characterized by multifractal modes.
Using modal analysis, we show that the quasiperiodic modulation reshapes the local density of states at the coupling site and thereby controls the effective spectral coupling between the emitter and the lattice.
The transient dynamics depend strongly on the spatial and spectral overlap between the emitter and the lattice modes: extended modes below the critical modulation generally produce rapid emitter depopulation, whereas critical and localized modes give rise to coherent back-action and persistent Rabi-like oscillations, with particularly pronounced behavior at the critical point.
We further identify emitter--photon bound states whose formation is controlled by the quasiperiodic spectral structure of the bath and whose overlap with the initially excited state leads to partial long-time population trapping.
Together, these results establish a direct connection between quasiperiodic localization, non-Markovian emitter dynamics, and emitter--photon bound-state formation.
In the weak-modulation regime, the long-time emitter population can therefore serve as a local spectroscopic probe of the AAH energy spectrum: enhanced population trapping identifies spectral gaps, whereas minimal or vanishing trapping reveals the spectral bands.

\end{abstract}

\maketitle

\section{\label{sec:level1}Introduction}

Coupling quantum emitters~(QEs) to structured photonic environments, such as cavities and waveguides, enables control over their spontaneous emission and coherent quantum dynamics~\cite{Walther_2006,RevModPhys.95.015002}.
Control over light--matter interactions underpins a broad range of nanophotonic and quantum-technological applications, including single-photon sources~\cite{doi:10.1126/science.290.5500.2282}, quantum sensing and metrology~\cite{RevModPhys.89.035002,Giovannetti2011}, quantum networks~\cite{Kimble2008}, quantum simulation~\cite{Bloch2012NatPhys,95pq-6r5g,PhysRevLett.131.226601}, and nonlinear quantum optics~\cite{Chang2014}, thereby motivating extensive theoretical and experimental research.
Within a quantum description, the influence of a structured photonic environment on a QE is encoded in the bath spectral density, which, for a fixed transition dipole, is determined by the photonic-mode frequencies and their local field amplitudes at the emitter position~\cite{NovotnyHecht2012}.
In particular, spontaneous-emission dynamics provide a local probe of the bath local density of state (LDOS): in the weak-coupling Markovian regime, an enhanced (suppressed) LDOS at the QE transition frequency accelerates (inhibits) radiative decay, whereas sharp spectral features can induce non-Markovian feedback and photonic band gaps can support emitter--photon bound states~\cite{PhysRevLett.64.2418,10.1063/5.0316541,dgby-6471,PhysRevA.93.033833}.
Beyond modifying radiative decay, structured reservoirs can retain memory of earlier emitter--field interactions and coherently return excitations to the emitter, giving rise to non-Markovian dynamics~\cite{Saeid_APL,8rw5-m26h, RevModPhys.95.015002, StewartPRR,Ferreiraprx,PhysRevA.108.053717, PhysRevResearch.7.013140, PhysRevLett.124.043603,
10.1063/5.0292425, PhysRevA.98.053608, Ferreiraprx}.
Despite extensive theoretical and experimental progress, the combined influence of the spectral and spatial structure of a photonic bath on the non-Markovian dynamics of a locally coupled QE remains incompletely understood.

Open-quantum-system approaches provide a natural framework for describing QEs interacting with structured photonic baths~\cite{BreuerPetruccione2002}.
In the hybridized regime, the coupled system can often be characterized by complex normal-mode frequencies: their real parts determine the dressed-mode energies and splittings, whereas their imaginary parts encode the corresponding linewidths and decay rates.
Although this picture is useful for describing light--matter hybridization, a description based on only a few normal modes does not generally capture how the frequency- and position-dependent bath LDOS governs the full QE dynamics.
For a locally coupled QE, the LDOS fixes the frequency dependence of the bath spectral density and, consequently, of the memory kernel governing the nonlocal-in-time dynamics.
Such reservoir-induced memory effects can generate oscillatory decay and population revivals and, in multi-emitter systems, long-range collective interactions~\cite{Rivera2020,Noda2007,RevModPhys.87.347,Saeid_APL,longhi2026,Gonzlez-Tudela2024,PhysRevA.105.053706,PhysRevResearch.2.013238}.
In this work, we show how spatial variations in the bath LDOS govern the spectral and temporal response of a QE, giving rise to position-dependent hybridization and non-Markovian dynamics.

A paradigmatic model of a quasiperiodically structured photonic bath is the one-dimensional Aubry--Andr\'ne--Harper~(AAH) lattice~\cite{Harper_1955,aubry1980,PhysRevLett.103.013901}.
In its standard form, the AAH model consists of a nearest-neighbor tight-binding lattice with quasiperiodically modulated onsite energies, whose modulation strength drives a transition from extended and localized regimes, with multifractal critical eigenmodes at the transition~\cite{Harper_1955,aubry1980,Li2023,PhysRevLett.103.013901,PhysRevB.23.6422,PhysRevLett.109.106402,PhysRevLett.123.034301}.
Unlike conventional Anderson localization generated by random disorder, localization in the AAH model originates from a deterministic quasiperiodic potential and is accompanied by a fragmented energy spectrum containing multiple subbands and gaps.
These features make the AAH lattice a tunable setting for disentangling the roles of spectral gaps, criticality, and localization in QE dynamics.
We therefore consider a two-level QE locally coupled to a one-dimensional AAH photonic lattice in the single-excitation sector.

Using a modal description that incorporates the full set of AAH
eigenmodes, we show that the global localization regime alone does not
determine the QE dynamics. Instead, the frequency- and
position-dependent LDOS selected jointly by the emitter transition
frequency and coupling position resultin  emitter--bath hybridization,
population decay, and non-Markovian backflow.
Across the localization transition, this produces dynamical behavior
ranging from rapid emitter depopulation into extended modes to enhanced
coherent back-action near the critical point and long-lived recurrent
oscillations in the localized regime.
We further identify isolated QE--photon bound states whose formation
and contribution to the infinite-time-averaged QE population are
controlled by the quasiperiodic bath spectrum, resulting in partial
excitation trapping.
Together, these results connect quasiperiodic localization physics
with non-Markovian quantum optics and establish the QE response as a
local spectroscopic probe of the bath band and gap structure.

\section{Model, mathematics and methods}
\subsection{Model and mathematics}
\begin{figure}
    \centering
    \includegraphics[width=0.76\linewidth]{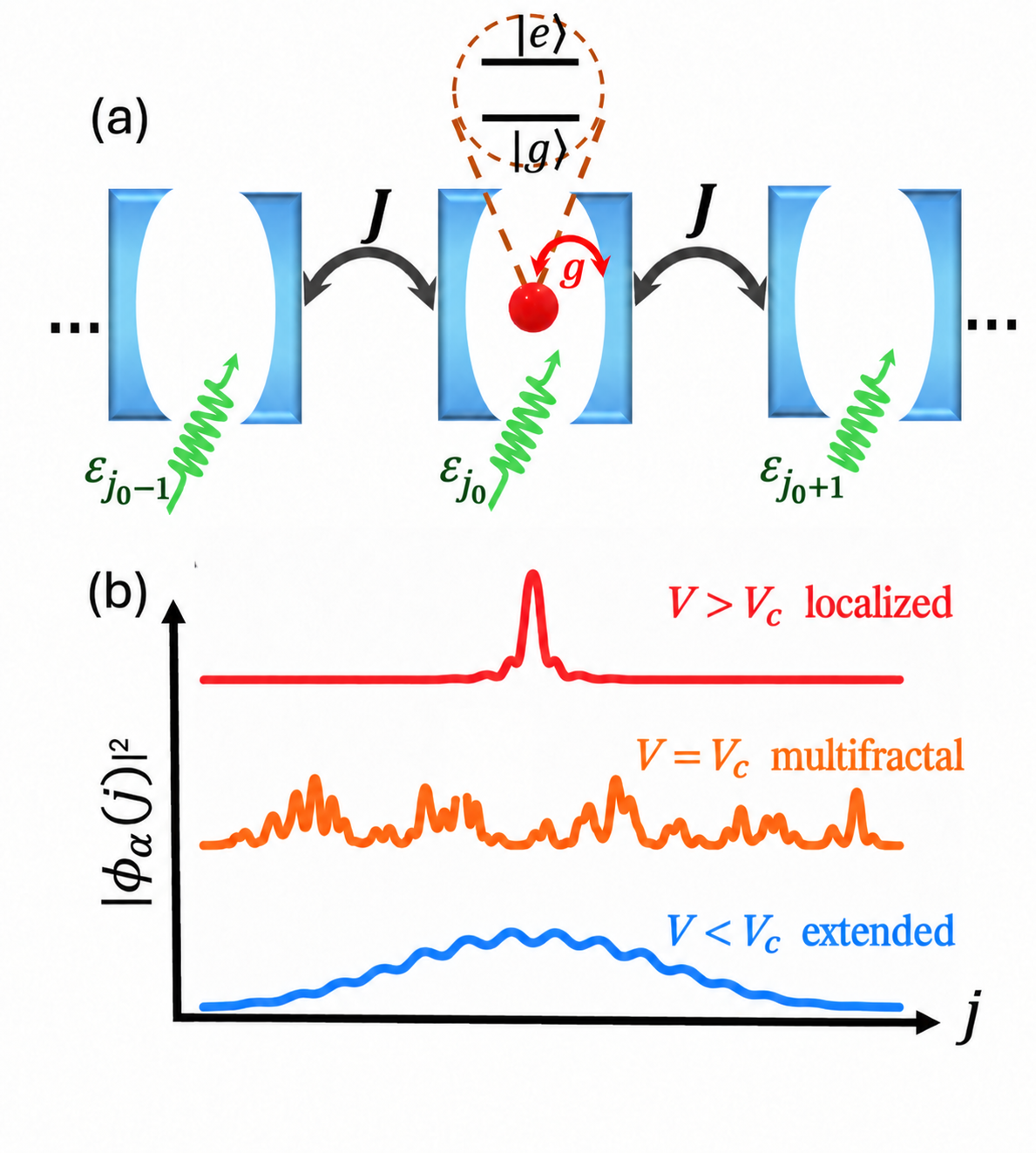}
    \caption{Schematic of a two-level QE coupled to an AAH photonic lattice. (a) Real-space representation of the system. The QE is locally coupled with strength $g$ to the cavity at site $j_0$. Photons tunnel between adjacent cavities with hopping amplitude $J$, while the onsite detunings in the frame rotating at $\omega_0$ follow the quasiperiodic modulation $\varepsilon_j=V\cos(2\pi\beta j+\phi)$.(b) Representative spatial intensity profiles $|\phi_\alpha(j)|^2$ of the AAH eigenmodes in the extended ($V<V_{\mathrm c}$), multifractal critical ($V=V_{\mathrm c}$), and localized ($V>V_{\mathrm c}$) regimes. In the eigenmode representation, the effective coupling of mode $\alpha$ to the QE is determined by its local amplitude at the emitter position, $g_\alpha=g\phi_\alpha(j_0)$. For the standard noninteracting AAH model, $V_{\mathrm c}=2J$ denotes the localization transition point. }
    \label{fig:Fig0.}
\end{figure}

\emph{Qualitative description}- We model the AAH lattice as a one-dimensional array of $N$ optically coupled cavities, each supporting a single mode with bare resonance frequency $\omega_0$.
Photons tunnel between adjacent cavities with nearest-neighbor hopping amplitude $J$.
A site-dependent external modulation shifts the resonance frequency of the $j$th cavity from \textbf{$\omega_0$ to $\omega_j = \omega_0 + \varepsilon_j$,} where $\varepsilon_j$ denotes the corresponding onsite-frequency shift.
Such onsite-frequency modulation can be implemented, for example, by varying the effective refractive index $n_{\mathrm{eff},j}$, the cavity length $L_j$, or both, according to the approximate scaling $\omega_j\propto c/(n_{\mathrm{eff},j}L_j)$.
A two-level QE, with ground~(excited) state $\ket{g}$~($\ket{e}$) and transition frequency $\omega_{\mathrm{eg}}$, is locally coupled to a selected cavity at site $j_0$, as illustrated in Fig.~\ref{fig:Fig0.}.
The primary aim of this work is to establish the spatially modulated LDOS of the AAH lattice as the central quantity describing QE--bath coupling, spontaneous-emission dynamics, and emitter--photon bound-state formation, rather than to develop a complete device-level description.
We therefore focus on the coherent single-excitation dynamics of the idealized system, neglecting cavity losses, emitter dephasing, and the multilevel structure of realistic emitters.

\emph{Quantitative description}- In the second-quantized tight-binding description, $b_j$~($b_j^\dagger$) denotes the annihilation~(creation) operator for a photon at the $j$th lattice site, while $\varepsilon_j$ denotes the corresponding onsite energy.
The quasiperiodic onsite modulation is defined as $\varepsilon_j=V\cos(2\pi\beta j+\phi)$,
where $V$ is the modulation strength, $\beta$ sets the incommensurability of the modulation with the underlying lattice, and $\phi$ is a global phase offset.
For irrational $\beta$, the standard AAH model exhibits a self-dual
localization transition at $V_{\mathrm c}=2J$ in the thermodynamic
limit~\cite{Harper_1955,aubry1980,PhysRevB.23.6422,PhysRevLett.103.013901,Li2023}. All single-particle eigenstates are extended for
$V<V_{\mathrm c}$, critical and multifractal at
$V=V_{\mathrm c}$, and exponentially localized for
$V>V_{\mathrm c}$. In the finite-size calculations, we approximate
the inverse golden ratio $\beta=(\sqrt{5}-1)/2$ by the consecutive
Fibonacci ratio $\beta=610/987$ and take $N=987$. Thus,
$V_{\mathrm c}=2J$ denotes the thermodynamic-limit critical value,
which is approached by the chosen finite approximant.

Working in a frame rotating at $\omega_0$, the AAH bath Hamiltonian in this rotating frame is then
\begin{equation}
H_{\mathrm{AAH}} =
\sum_{j=1}^{N} \varepsilon_j \, b_j^\dagger b_j - J \sum_{j=1}^{N} \left(
b_{j+1}^\dagger b_j + b_j^\dagger b_{j+1}
\right).
\label{eq:Hbath}
\end{equation}

The lowering and raising operators of the two-level QE are defined as
$\sigma:=\ket{g}\bra{e}$ and $\sigma^\dagger:=\ket{e}\bra{g}$, respectively.

Setting $\hbar=1$, the QE Hamiltonian in the frame rotating at $\omega_0$ is
\begin{equation}
H_{\mathrm{QE}} = \Delta_{\mathrm{eg}}\sigma^\dagger\sigma, \qquad \Delta_{\mathrm{eg}} = \omega_{\mathrm{eg}}-\omega_0,
\label{eq:HQE}
\end{equation}

We then introduce the quantum mechanical dipole-moment operator of this QE as $\hat{\mathbf{p}}=(\mathbf{p}_{ge} \sigma + \mathbf{p}_{ge}^* \sigma^\dagger)$,
where
$\mathbf{p}_{ge}=\bra{g}\hat{\mathbf{p}}\ket{e}$ is the transition dipole moment. Assuming a real transition dipole, this reduces to
$\hat{\mathbf{p}}=\mathbf{p}_{ge}(\sigma+\sigma^\dagger)$.
Next, we use the dipole approximation $H_\text{int}=-\hat{\mathbf{p}}\cdot\hat{\mathbf{E}}(\bm{j}_{0})$; $\bm{j}_{0}$ the position of the QE and the quantized electric field as $\hat{\mathbf{E}}(\bm{j}_0)\propto(b_{j_0}+b_{j_0}^\dagger)$, to describe the interaction between QE and AAH lattice within the rotating-wave approximation. 
The interaction Hamiltonian becomes
\begin{equation}
H_{\mathrm{int}} = g\left( b_{j_0}^\dagger\sigma+b_{j_0}\sigma^\dagger \right),
\label{eq:Hint}
\end{equation}
where $g$ is the local QE--cavity coupling strength.
The QE--AAH system is illustrated schematically in Fig.~\ref{fig:Fig0.}.
Combining Eqs.~\eqref{eq:Hbath}--\eqref{eq:Hint}, the total Hamiltonian in the frame rotating at $\omega_0$ is
\begin{equation}
H=H_{\mathrm{QE}}+H_{\mathrm{AAH}}+H_{\mathrm{int}}.
\label{Eq:Htotal}
\end{equation}
Within the rotating-wave approximation, the total excitation number is conserved. We therefore restrict the dynamics to the single-excitation sector spanned by
$\{\ket{e,0},\ket{g,1_j}\}$, where
$\ket{e,0}:=\ket{e}\otimes\ket{0}$ denotes the excited QE with the lattice in its vacuum state, and
$\ket{g,1_j}:=\ket{g}\otimes\ket{1_j}$ denotes the QE in its ground state with one photon at lattice site $j$, with $\ket{1_j}=b_j^\dagger\ket{0}$.
The state of the combined system can then be written as
\begin{equation}
|\psi(t)\rangle = c_{\mathrm e}(t) \ket{e,0} + \sum_{j=1}^{N} c_j(t)\ket{g,1_j}.
\label{Eq:Wavefunction}
\end{equation}
Here, $c_{\mathrm e}(t)$ is the probability amplitude for the excitation to reside in the QE, whereas $c_j(t)$ is the probability amplitude for finding the photon at lattice site $j$. Accordingly, $|c_{\mathrm e}(t)|^2$ gives the excited-state population of the QE, while $|c_j(t)|^2$ gives the single-photon population at site $j$, with $|c_{\mathrm e}(t)|^2+\sum_{j=1}^{N}|c_j(t)|^2=1$.

\subsection{Methods}
The real-space representation of $H_{\mathrm{AAH}}$ in Eq.~\eqref{eq:Hbath} is convenient for describing photon hopping through the lattice, but it does not explicitly resolve the contributions of individual bath eigenmodes to the QE dynamics.
To expose this spectral structure, we diagonalize the AAH Hamiltonian according to
$ H_{\mathrm{AAH}} \ket{\alpha} = \omega_\alpha \ket{\alpha}$,  
where $\ket{\alpha}$ denotes the $\alpha$th single-photon eigenmode and $\omega_\alpha$ is its eigenfrequency measured relative to $\omega_0$ in the rotating frame.
The site-basis operators are related to the eigenmode operators through the unitary transformation
\begin{equation}
b_j=\sum_{\alpha=1}^{N}\phi_{\alpha j}a_\alpha,
\qquad
b_j^\dagger=\sum_{\alpha=1}^{N}\phi_{\alpha j}^{*}a_\alpha^\dagger, 
\label{Eq:modeTransformation}
\end{equation}

Here, $a_\alpha$ annihilates a photon in the $\alpha$th AAH eigenmode, whose expansion in the site basis is
$\ket{\alpha} = \sum_{j=1}^{N}\phi_{\alpha j}\ket{1_j},
\phi_{\alpha j}:=\braket{1_j|\alpha}$, 
with the normalization $\sum_{j=1}^{N}|\phi_{\alpha j}|^2=1$.
In the eigenmode basis, a QE locally coupled to site $j_0$ interacts with eigenmode $\alpha$ through the coupling amplitude $ g_\alpha:=g\phi_{\alpha n_0}$,
whose magnitude is determined by the local mode weight at the emitter position, $|g_\alpha|^2=g^2|\phi_{\alpha n_0}|^2$.
We characterize the QE--bath coupling through the bath LDOS and obtain the QE dynamics by solving the Schr"odinger equation in the eigenmode basis.
The bath LDOS at lattice site $j$ is defined as
\begin{equation}
\rho_j(\omega) = \sum_{\alpha=1}^{N} |\phi_{\alpha j}|^2 \delta(\omega-\omega_\alpha).
\label{Eq:LDOS}
\end{equation}
It quantifies the frequency-resolved local spectral weight of the bath modes. For a QE coupled to site $j_0$, the corresponding bath spectral density is
$\mathcal{J}_{n_0}(\omega) =  g^2\rho_{j_0}(\omega)$.
Consequently, spatial variations in $\rho_j(\omega)$ reveal how the spectral weight at frequency $\omega$ is distributed across the lattice, while $\rho_{j_0}(\omega)$ determines which bath modes are locally accessible to the QE.

Using Eq.~\eqref{Eq:modeTransformation} to express the single-excitation wavefunction in the AAH eigenmode basis and eliminating the bath-mode amplitudes from the Schr"odinger equation, we obtain the following equation for the QE excited-state amplitude (see Appendix~\ref{ap:kernel} for more details):
\begin{equation}
\dot{c}_{\mathrm e}(t) = -i\Delta_{\mathrm{eg}}c_{\mathrm e}(t)
-\int_0^t d\tau K_{j_0}(t-\tau)c_{\mathrm e}(\tau),
\label{TLSdyn.eqn}
\end{equation}
where the position-dependent memory kernel is
\begin{equation}
K_{j_0}(t-\tau) = \int_{-\infty}^{\infty}d\omega \mathcal{J}_{j_0}(\omega)e^{-i\omega(t-\tau)}.
\label{Kernel.eqn}
\end{equation}
Thus, the memory kernel is the Fourier transform of the local spectral density or, equivalently, $g^2$ times the Fourier transform of the LDOS sampled at the emitter position. This relation directly connects the local spectral structure of the AAH bath to the feedback experienced by the QE.
If $\mathcal{J}_{j_0}(\omega)$ is broad and slowly varying over the frequency range relevant to the QE, the memory kernel decays on a timescale much shorter than that of the QE dynamics, permitting a Markovian description with approximately exponential decay. By contrast, pronounced spectral features, such as narrow peaks, band edges, gaps, or fragmented minibands, can generate a long-lived memory kernel and non-Markovian dynamics, including population revivals and damped Rabi-like oscillations. Moreover, because $\mathcal{J}_{j_0}(\omega)=g^2\rho_{j_0}(\omega)$, spatial variations in the LDOS render the QE dynamics sensitive to the coupling position $j_0$.

\begin{figure}[t]
    \centering
    \includegraphics[width=1
    \linewidth]{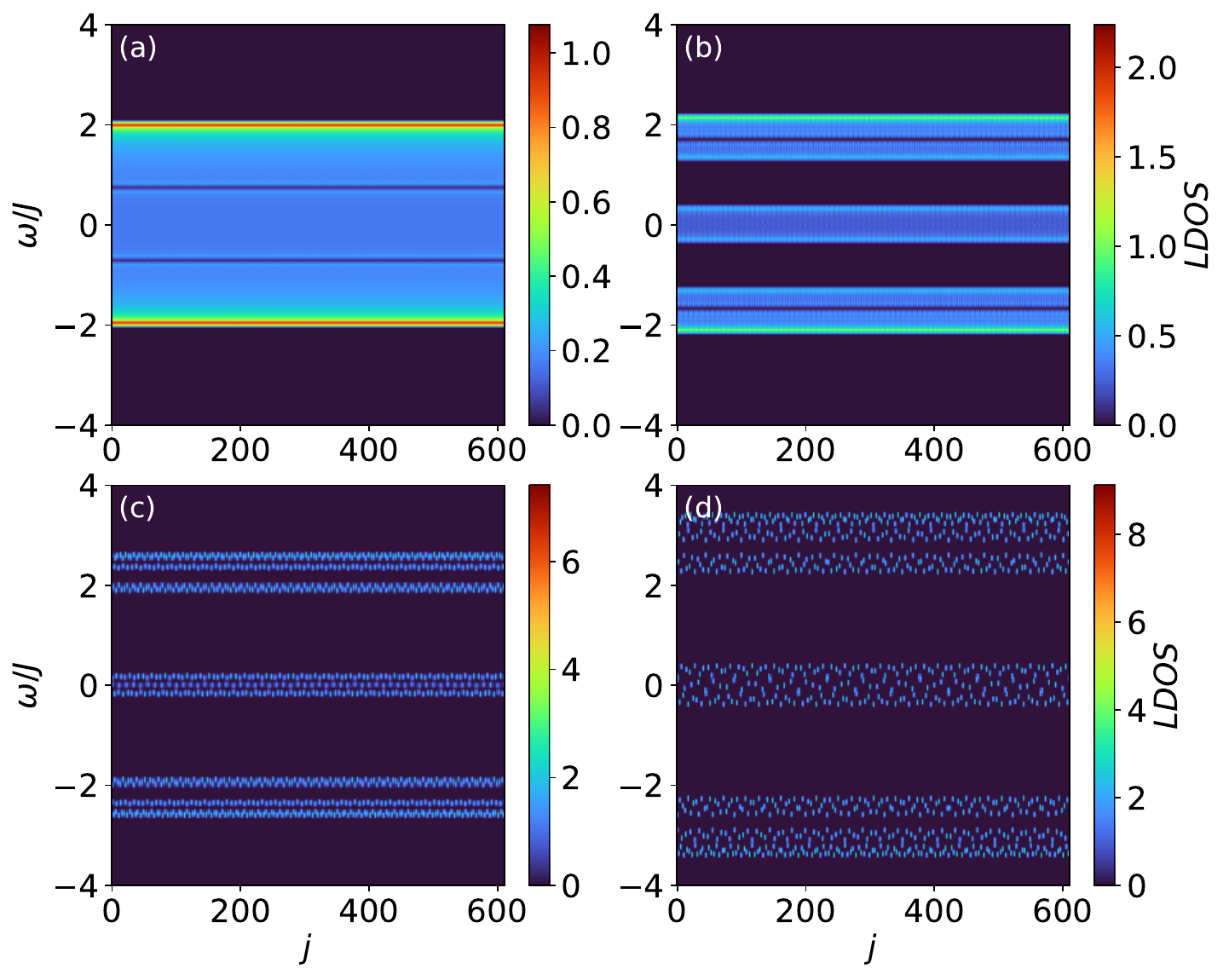}
    \caption{Spatially resolved LDOS $\rho_j(\omega)$ of the AAH lattice as a function of normalized frequency $\omega/J$ and lattice position $j$. Panels~(a)--(d) correspond to modulation strengths $V/J=0.1$, $1$, $2$, and $3$, respectively. The color scale represents the LDOS magnitude. The LDOS is evaluated on a uniform frequency grid with spacing $d\omega=0.001J$ using a frequency-bin width $\Delta\omega=0.1J$. At every lattice site, it satisfies the sum rule $\int_{-\infty}^{\infty}\rho_j(\omega),d\omega=1$. The remaining parameters are $N=987$, $\beta=610/987$, and $\theta=0$. }
    \label{Fig2}
\end{figure}

\section{Results}

\subsection{Spectral and spatial structure of the AAH bath}

Because the memory kernel in Eq.~\eqref{Kernel.eqn} is determined by the LDOS at the emitter position, we first examine how the AAH localization transition reshapes the frequency- and position-dependent spectral structure of the bath. The extended, critical, and localized regimes considered below are independently confirmed through the IPR analysis presented in Appendix~\ref{ap:IPR}.
Figure~\ref{Fig2} shows the spatially resolved LDOS $\rho_j(\omega)$ as a function of frequency $\omega$ and lattice position $j$ for representative values of the modulation strength $V$. 
In the extended regime, represented by $V=0.1J$ and $V=J$ in Figs.~\ref{Fig2}(a) and \ref{Fig2}(b), respectively, the LDOS is broadly distributed across the lattice and exhibits comparatively weak spatial variations, consistent with the extended character of the bath eigenmodes.
Consequently, the local spectral density sampled by the QE varies only weakly with the coupling position $j_0$, and the resulting QE dynamics are expected to be comparatively insensitive to position in this regime.
Within the extended regime, the LDOS is enhanced near the band edges, while increasing the modulation strength from $V=0.1J$ to $V=J$ progressively restructures the bath spectrum and redistributes the local spectral weight over frequency.

The spatial and spectral structure of the LDOS changes qualitatively at the critical point $V=V_{\mathrm c}=2J$ and in the localized regime represented by $V=3J$.
At the critical point, Fig.~\ref{Fig2}(c) shows that the LDOS develops pronounced spatial fluctuations and a fragmented spectral structure.
These features are consistent with the known multifractal character
of the critical AAH eigenmodes, while the IPR analysis in Appendix~\ref{ap:IPR}
provides a complementary qualitative characterization of the
localization regimes.
In the localized regime at $V=3J$, Fig.~\ref{Fig2}(d) shows that the LDOS becomes concentrated at specific lattice positions and frequencies, reflecting the exponential localization of the bath eigenmodes.
Consequently, at and beyond the localization transition, the local spectral density $\mathcal{J}_{j_0}(\omega)=g^2\rho_{j_0}(\omega)$ becomes strongly dependent on both frequency and coupling position. Through Eq.~\eqref{Kernel.eqn}, this structure produces a position-dependent memory kernel and provides the basis for understanding the QE dynamics analyzed below.
\begin{figure}[t]
    \centering
    \includegraphics[width=1
    \linewidth]{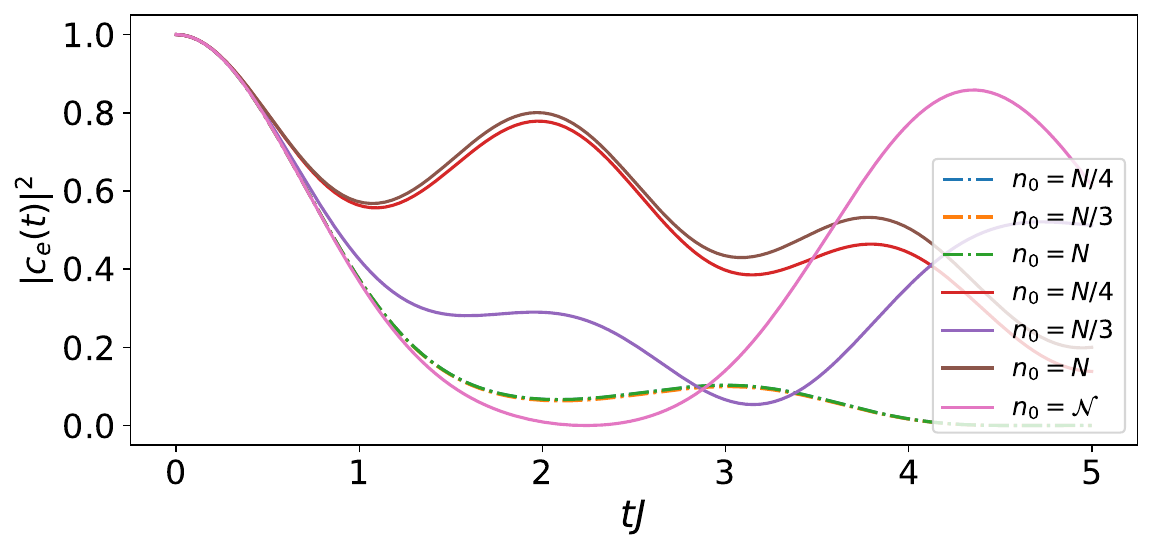}
    \caption{QE excited-state population dynamics $|c_{\mathrm e}(t)|^2$ of QE for different coupling positions in the AAH lattice in (a) and localized LDOS in (b). The QE is coupled at positions $n_0 =  [N/4],\, [N/3],\, N$ and $\mathcal N = 341$. Two values of the quasiperiodic modulation strength are considered: $V/J = 0.1$ (dot-dashed lines) and $V/J = 2$ (solid lines).  
    The coupling strength $g/J = 1$, emitter transition frequency $\Delta_{\mathrm{eg}} = 0$, the brackets $[\cdot]$ denotes the integer part,  and rest of parameters is same as Fig.~\ref{Fig2}.}
    \label{Fig3}
\end{figure}
\subsection{Position-dependent QE dynamics}

Having characterized the spatially resolved LDOS, we now examine how the modulation strength $V$ and coupling position $j_0$ govern the excited-state population $P_{\mathrm e}(t)=|c_{\mathrm e}(t)|^2$ for a QE resonant with the bare cavity frequency, $\omega_{\mathrm{eg}}=\omega_0$, corresponding to $\Delta_{\mathrm{eg}}=0$ in the rotating frame.
Figure~\ref{Fig3} compares the QE dynamics in two representative regimes: the extended regime, $V<V_{\mathrm c}$, and the critical regime, $V=V_{\mathrm c}=2J$, where the bath eigenmodes are multifractal.
In the extended regime, each normalized bath eigenmode is distributed over the entire lattice, such that its characteristic local spectral weight scales as $|\phi_{\alpha j_0}|^2\sim 1/N$.
Consequently, the LDOS varies only weakly among the coupling positions considered, as illustrated in Figs.~\ref{Fig2}(a) and \ref{Fig2}(b).
Consistent with this weak spatial variation, the dash-dotted curves in Fig.~\ref{Fig3} nearly coincide for $j_0=\lfloor N/4\rfloor$, $N/3$, and $N$, demonstrating that the QE population dynamics depend only weakly on the coupling position in the extended regime.
Around the emitter resonance, $\Delta_{\mathrm{eg}}=0$, the comparatively smooth LDOS produces a short-lived memory kernel (Markovian dynamics) and an approximately monotonic initial decay of the QE population over the time interval shown.
Because the lattice is finite, however, the emitted excitation can eventually revisit the coupling site and produce a finite-size recurrence at time $T_{\mathrm{rec}}$, whose dependence on the lattice size is analyzed in Fig.~\ref{FigS3p}.
At the critical point, $V=V_{\mathrm c}$, the multifractal bath character of the bath modes and the fragmented spectrum produce pronounced site-to-site variations in the LDOS, as shown in Fig.~\ref{Fig2}(c).
Consequently, the solid curves in Fig.~\ref{Fig3} exhibit remarkably different time evolutions for different values of $j_0$, demonstrating the strong position dependence of the QE dynamics at criticality.
When the QE couples to a site whose LDOS exhibits a pronounced and spectrally narrow feature near the resonance frequency, the resulting long-lived memory kernel enhances bath back-action and gives rise to non-Markovian dynamics.
For example, at $j_0=\mathcal{N}$, with $\mathcal{N}=341$, where the resonant LDOS is maximal, the QE population undergoes near-complete depletion followed by a pronounced revival, as shown by the pink curve in Fig.~\ref{Fig3}.

To clarify the physical origin of these contrasting QE dynamics, we examine the spatial propagation of the emitted excitation in Appendix~\ref{ap:recurrence}. In the weak-modulation regime, $V\ll V_{\mathrm c}$, the excitation spreads ballistically and returns to the coupling site after traversing the finite periodic lattice, producing a recurrence time that scales approximately linearly with $N$. At the critical point, multifractal bath modes strongly inhibit ballistic spreading and enhance the local return probability, producing population revivals without requiring propagation around the entire lattice.
Together, these results show that the AAH transition changes the QE response from nearly position-insensitive pre-recurrence dynamics in the extended regime to strongly position-dependent non-Markovian dynamics governed by the local spectral structure of the critical bath.

\begin{figure}[t]
    \centering
    \includegraphics[width=1
    \linewidth]{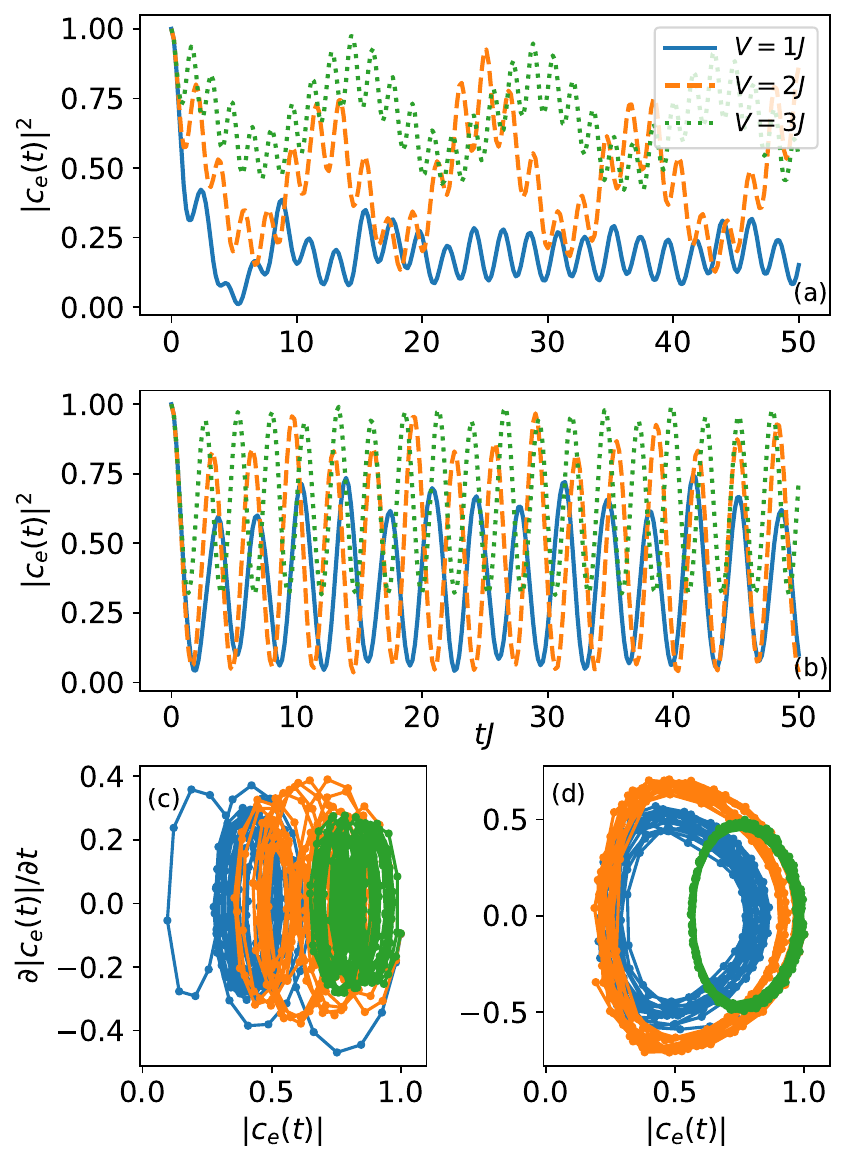}
    \caption{QE excited-state dynamics and corresponding phase-plane projections for different AAH modulation strengths and emitter detunings.
    Panels~(a) and (b) show the excited-state population $|c_{\mathrm e}(t)|^2$ for $\Delta_{\mathrm{eg}}=0$ and $\Delta_{\mathrm{eg}}=2J$, respectively.
    The solid blue, dashed orange, and dotted green curves correspond to $V/J=1$, $2$, and $3$, respectively.
    Panels~(c) and (d) display the corresponding phase-plane projections $\bigl(|c_{\mathrm e}(t)|,\partial_t|c_{\mathrm e}(t)|\bigr)$ for the dynamics in panels~(a) and (b), respectively, using the same line styles and color coding. The calculations use coupling position $j_0=\lfloor N/4\rfloor$ under periodic boundary conditions.}
    \label{Fig4}
\end{figure}

\subsection{Spectral dependence of the QE dynamics}
Having established how the modulation strength and coupling position affect the QE dynamics, we now examine how tuning the emitter detuning $\Delta_{\mathrm{eg}}$ across the bath spectrum modifies its emission dynamics.
We first consider the extended regime, $V=J<V_{\mathrm c}$, and set $\Delta_{\mathrm{eg}}=0$ and assume QE being locally coupled at $j_0=\lfloor N/4\rfloor$. 
At this coupling site, the comparatively smooth LDOS around $\omega=0$ in Fig.~\ref{Fig2}(b) enables the QE to couple to a broad set of extended bath modes, producing near-complete initial depletion of the excited-state population.
At later times, the population remains small and exhibits only weak residual oscillations, as shown by the solid blue curve in Fig.~\ref{Fig4}(a).
Increasing the modulation strength to $V=V_{\mathrm c}$ and subsequently to $V=3J$, represented by the dashed orange and dotted green curves in Fig.~\ref{Fig4}(a), respectively, leads to an increasingly fragmented LDOS with pronounced spectral features near the QE resonance [Fig.~\ref{Fig2}(c) and (d)].

This spectral restructuring concentrates the local spectral weight within a restricted set of bath modes, thereby enhancing coherent QE--bath exchange and producing the population revivals observed in Fig.~\ref{Fig4}(a).
The nonvanishing population over the displayed time interval further indicates a slowly decaying component, whose connection to emitter--photon bound states is examined below.
For $V=3J>V_{\mathrm c}$, the bath eigenmodes are localized, restricting the QE interaction predominantly to modes with appreciable amplitude at the coupling site $j_0$.
Consequently, the emitted excitation remains concentrated near the coupling region, enabling repeated coherent exchange with the QE and producing the pronounced population oscillations shown by the dotted green curve.
We next set $\Delta_{\mathrm{eg}}=2J$. In contrast to the $\Delta_{\mathrm{eg}}=0$, the QE does not relax toward a low excited-state population; instead, its dynamics are dominated by persistent, large-amplitude oscillations, as shown in Fig.~\ref{Fig4}(b).
These sustained oscillations are consistent with the strongly structured LDOS near $\omega=\Delta_{\mathrm{eg}}=2J$, which concentrates the local spectral weight within a restricted set of near-resonant bath modes rather than a broad continuum
A comparison of the three curves in Fig.~\ref{Fig4}(b) shows that both the oscillation period and amplitude depend on $V$: the critical and localized regimes exhibit nearly complete population revivals, whereas the revival amplitude remains smaller for $V=J$ and higher at critical point.
This behavior is inherently position dependent because changing $j_0$ modifies the mode-resolved couplings $g_\alpha=g\phi_{\alpha j_0}$ and therefore the set of bath modes participating in the QE dynamics.

The phase-plane projections $\bigl(|c_{\mathrm e}(t)|,\partial_t|c_{\mathrm e}(t)|\bigr)$ in Figs.~\ref{Fig4}(c) and \ref{Fig4}(d) provide a complementary visualization of the recurrent QE dynamics. For $\Delta_{\mathrm{eg}}=0$, the projections form comparatively dense and irregular trajectories [Fig.~\ref{Fig4}(c)], consistent with interference among multiple bath-induced frequency components. 
For $\Delta_{\mathrm{eg}}=2J$, by contrast, they form more regular and well-defined loop-like structures [Fig.~\ref{Fig4}(d)], consistent with dynamics governed by a smaller number of dominant coherent frequency components selected by the structured spectral-edge LDOS. 
In Fig.~\ref{Fig4}(d), the critical trajectory spans the largest range of $\partial_t|c_{\mathrm e}(t)|$, reflecting the comparatively rapid variation of the emitter amplitude and the strong bath feedback at $V=V_{\mathrm c}$.
These phase-plane projections provide only a qualitative representation of recurrence and should not be interpreted as an independent quantitative criterion for non-Markovianity. We therefore employ the normalized population-backflow measure $\chi\in[0,1]$, whose formal definition is given in Appendix~\ref{Decay}. Here, $\chi=0$ indicates that no population backflow is detected over the observation interval, whereas values approaching unity indicate increasingly strong non-Markovian backflow.

The results in Fig.~\ref{FigS4} show that, for $\Delta_{\mathrm{eg}}=0$, $\chi$ increases from approximately $0.55$ at $V=J$ to $0.90$ at $V=V_{\mathrm c}$, before decreasing slightly to $0.85$ at $V=3J$. Thus, for the selected coupling site and observation interval, the population backflow is strongest at the critical point. For $\Delta_{\mathrm{eg}}=2J$, the repeated population revivals produce substantial backflow for all three modulation strengths, including the extended regime at $V=J$. Together, these results show that spectral-edge tuning supports robust recurrent dynamics even in the extended regime, whereas at the band center the non-Markovian response is strongly enhanced by the critical bath structure.

\subsection{QE--photon bound states and excitation trapping}

The persistent oscillations and nonzero QE population observed over the finite time interval in Fig.~\ref{Fig4} indicate the presence of long-lived dressed components, but do not by themselves establish bound-state formation. We therefore analyze the isolated eigenenergies of the coupled QE--AAH system to determine whether these dynamical components can originate from QE--photon bound states.
In the thermodynamic limit, a bound state is a normalizable eigenstate of the coupled Hamiltonian whose real energy lies outside the continuous bath spectrum or within a genuine spectral gap, with a photonic component localized around the emitter. The quasiperiodic modulation fragments the AAH spectrum into minibands separated by spectral gaps, thereby providing additional energy windows bound states formation.
Applying a Laplace-transform analysis to the single-excitation equations yields the pole condition for the dressed eigenenergies $E$~\cite{Gonzalez-Tudela2017,PhysRevLett.64.2418},
\begin{equation}
E-\Delta_{\mathrm{eg}} = \Sigma_{j_0}(E),
\qquad \Sigma_{j_0}(E) = g^2\sum_{\alpha=1}^{N}
\frac{|\phi_{\alpha j_0}|^2}{E-\omega_\alpha}.
\label{SS.eqn}
\end{equation}
In the continuum limit, the corresponding retarded self-energy is
\begin{equation}
\Sigma_{j_0}^{R}(E) = \mathcal{P}\!\int d\omega\, \frac{\mathcal{J}_{j_0}(\omega)}{E-\omega} -i\pi\mathcal{J}_{j_0}(E).
\nonumber
\label{Eq:retardedSelfEnergy}
\end{equation}
Its real part gives the bath-induced level shift, often referred to as a Lamb shift, whereas
$\Gamma_{j_0}(E)=-2\,\operatorname{Im}\Sigma_{j_0}^{R}(E)
=2\pi\mathcal{J}_{j_0}(E)$
gives the decay rate into the available bath modes.
For finite $N$, Eq.~\eqref{SS.eqn} yields both bath-like dressed solutions continuously connected to the eigenmodes of the uncoupled AAH lattice and additional isolated solutions induced by the QE--bath interaction. In the following, we retain only the latter solutions whose energies lie outside the bath spectral support or within genuine spectral gaps. In these regions, $\mathcal{J}_{j_0}(E_b)=0$ and hence $\operatorname{Im}\Sigma_{j_0}^{R}(E_b)=0$, allowing the corresponding solutions to be identified as bound states.
\begin{figure}[t]
    \centering
    \includegraphics[width=\linewidth]{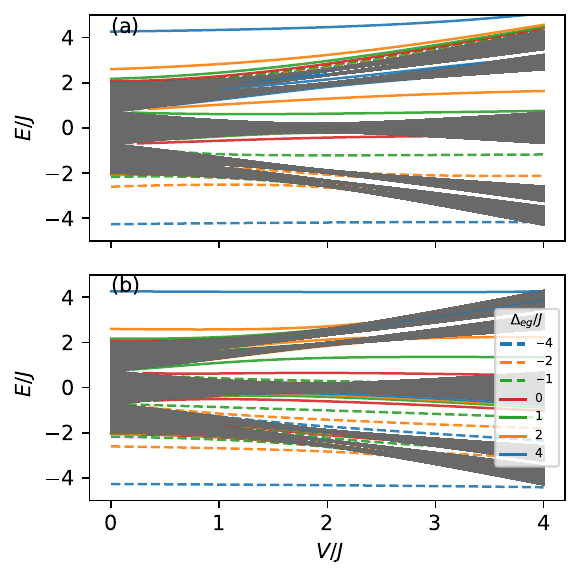}
    \caption{\label{Fig5}
    QE--photon bound-state energies as functions of the modulation strength $V/J$.
    Panels~(a) and (b) correspond to coupling positions $j_0=\lfloor N/4\rfloor$ and $j_0=\lfloor N/2\rfloor$, respectively.
    The blue, orange, green, and red curves correspond to $|\Delta_{\mathrm{eg}}|/J=4$, $2$, $1$, and $0$, respectively; dashed and solid curves denote negative and positive detunings.
    The gray regions show the spectrum of the noninteracting AAH bath.
    The remaining parameters are the same as in Fig.~\ref{Fig3}.
    }
\end{figure}
Figure~\ref{Fig5} shows how QE--photon bound states are  controlled by the modulation strength, emitter detuning $\Delta_{\mathrm {eg}}$, and coupling position. For a fixed detuning, multiple bound-state branches can coexist once the quasiperiodic modulation opens additional spectral gaps. One branch generally remains close to the detuning energy, while additional branches emerge from the fragmented gap structure of the bath.
A comparison of panels~(a) and (b) shows that changing $j_0$ shifts the bound-state branches and can alter their number even when $V$ and $\Delta_{\mathrm{eg}}$ remain fixed. This position dependence follows directly from the self-energy in Eq.~\eqref{SS.eqn}, because different coupling sites sample different mode weights $|\phi_{\alpha j_0}|^2$ and therefore different local spectral environments.
In the weak-modulation regime, $V\ll V_{\mathrm c}$, the bath spectrum is only weakly fragmented, and the isolated bound-state solutions occur predominantly outside the outer band edges. Negative and positive detunings favor branches below and above the bath spectrum, respectively, while increasing $|\Delta_{\mathrm{eg}}|$ shifts the corresponding branch farther from the bath spectral.
As $V$ increases, the formation and reorganization of miniband gaps cause additional bound-state branches to emerge and shift across the spectrum. Simultaneously, increasing the modulation strength can cause some previously isolated branches to merge into the bath bands and cease to satisfy the bound-state criterion. Thus, quasiperiodicity provides a tunable mechanism for controlling both the number and energies of bound states through the combined effects of spectral fragmentation and position-dependent local mode weights.

Having identified the isolated bound-state branches, we now quantify their dynamical consequence through the QE population retained by these states.
Because bound-state components do not decay into propagating bath modes, any overlap of the initial state with the bound-state subspace produces a persistent contribution to the QE excited-state population.
For nondegenerate bound-state energies, the infinite-time-averaged QE population contributed by this subspace is
$
P_{\mathrm{QE}}^{\mathrm{BS}}
=
\sum_{\mu\in\mathrm{BS}}
\left|\langle\psi_\mu|\psi_0\rangle\right|^4.
$
Here, $\ket{\psi_0}=\ket{e,0}$ is the initial state, $\ket{\psi_\mu}$ denotes a normalized  bound-state eigenstate, and the sum runs over all bound states. which can be rewritten as
\begin{equation}
P_{\mathrm{QE}}^{\mathrm{BS}} = 
\sum_{\mu\in\mathrm{BS}}\left[
1+g^2\sum_{\alpha}
\frac{|\phi_{\alpha j_0}|^2}
{(E_{\mu}-\omega_\alpha)^2}
\right]^{-2}.
\label{Eq:residue}
\end{equation}

\begin{figure}[t]
    \centering
\includegraphics[width=\linewidth]{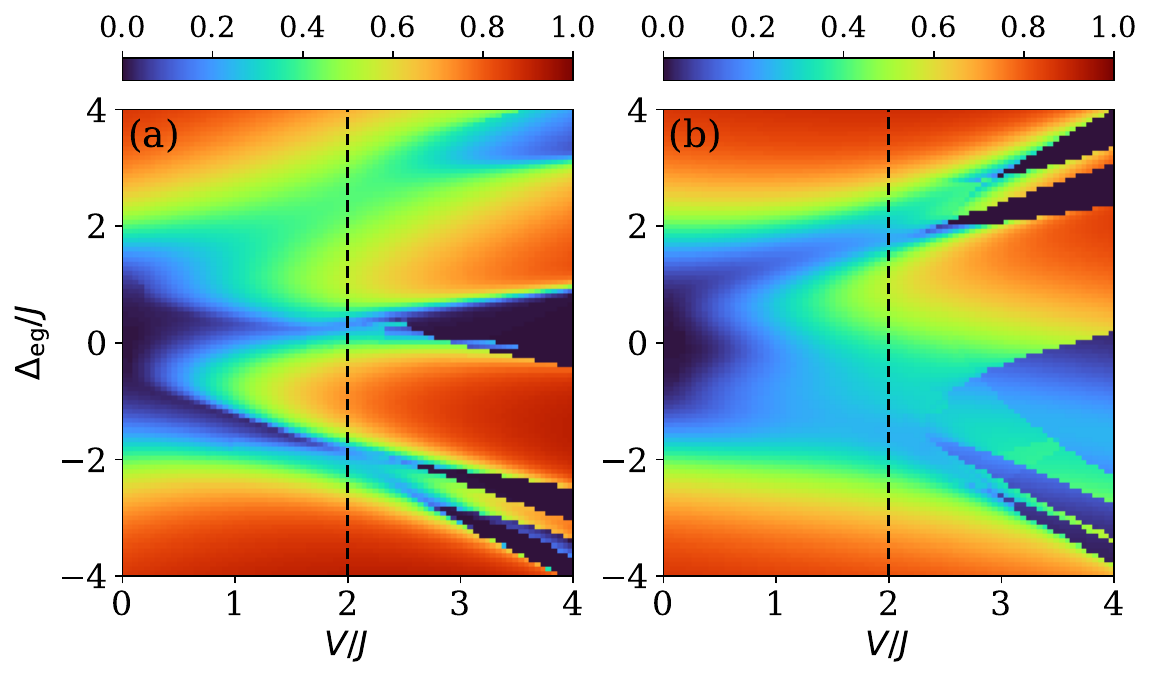}
    \caption{\label{Fig6}
    Infinite-time-averaged bound-state contribution to the QE population, $P_{\mathrm{QE}}^{\mathrm{BS}}$, as a function of the emitter detuning $\Delta_{\mathrm{eg}}/J$ and modulation strength $V/J$.
    Panels~(a) and (b) correspond to the coupling positions $j_0=\lfloor N/4\rfloor$ and $j_0=\lfloor N/2\rfloor$, respectively.
    The vertical dashed line marks the localization transition at $V=V_{\mathrm c}$.
    }
\end{figure}

For finite $N$, all $N+1$ solutions of Eq.~\eqref{SS.eqn} are discrete dressed eigenenergies, most of which remain continuously connected to the uncoupled bath modes. A solution cannot therefore be identified as a bound state solely because it differs from a bare bath eigenenergy. We retain as candidate bound states only those solutions lying outside the outer bath edges or within resolved miniband gaps. Their bound-state character is further established by verifying that the corresponding gap and eigenenergy persist with increasing system size and that the photonic component remains localized around the QE position.
Figure~\ref{Fig6} shows the infinite-time-averaged QE population contributed by the bound-state subspace, $P_{\mathrm{QE}}^{\mathrm{BS}}$, as a function of the modulation strength $V/J$ and emitter detuning $\Delta_{\mathrm{eg}}/J$ for the coupling positions $j_0=\lfloor N/4\rfloor$ and $j_0=\lfloor N/2\rfloor$.
The resulting density plots show that the bound-state contribution to excitation trapping is strongly nonuniform across the $(V,\Delta_{\mathrm{eg}})$ parameter space and varies substantially with the coupling position $j_0$.

In the extended regime, $V<V_{\mathrm c}$, enhanced $P_{\mathrm{QE}}^{\mathrm{BS}}$ occurs predominantly for emitter detunings outside or near the outer edges of the bath spectrum, consistent with the formation of isolated out-of-band bound states. This behavior suggests that the bound-state contribution to the QE population can serve as a local probe of the bath band edges and spectral gaps.
For $j_0=\lfloor N/4\rfloor$ shown in Fig.~\ref{Fig6}(a), the region of enhanced $P_{\mathrm{QE}}^{\mathrm{BS}}$ broadens toward positive detunings as $V$ increases, consistent with the emergence of multiple bound-state branches within the quasiperiodicity-induced gaps. By contrast, for $j_0=\lfloor N/2\rfloor$ in panel~(b), this region broadens predominantly toward negative detunings. Measurements at different coupling positions therefore provide complementary information about the position-dependent local spectral structure of the bath.
As $V$ approaches $V_{\mathrm c}$, the increasing fragmentation of the bath spectrum and the opening of miniband gaps produce narrow ridges of nonzero $P_{\mathrm e}^{\mathrm{BS}}$, reflecting the formation of gap-bound states.
In the localized regime, $V>V_{\mathrm c}$, the map of $P_{\mathrm QE}^{\mathrm{BS}}$ becomes highly irregular and position dependent because the QE couples appreciably only to a restricted set of bath modes localized near the coupling site.
Consequently, a map of $P_{\mathrm{QE}}^{\mathrm{BS}}$ would not accurately capture the local spectral environment and the miniband structure.
Taken together, these results show that scanning the QE detuning and coupling position enables bound-state-assisted spectroscopy of the quasiperiodic bath: the global gap structure is most clearly resolved in the extended regime, whereas the local spectral environment dominates in the localized regime.

\section{summary and conclusion}
In this work, we investigated the spontaneous-emission dynamics of a QE locally coupled to a one-dimensional quasiperiodic AAH photonic lattice within the single-excitation sector.
By expressing the QE--bath interaction in the bath eigenmode basis, we showed that the site-resolved LDOS determines the effective QE--bath coupling, the non-Markovian memory kernel, and the resulting emission dynamics.
In the extended regime, $V<V_{\mathrm c}$, the bath eigenstates are delocalized, and the LDOS remains comparatively uniform across the lattice.
As a result, the QE dynamics are largely insensitive to the coupling position, and the excitation decays efficiently into extended bath modes with relatively weak memory effects
Near the localization transition, $V\simeq V_{\mathrm c}$, the fragmentation of the bath spectrum into minibands becomes pronounced, while the LDOS develops strong spatial inhomogeneity.
The multifractal character of the critical modes produces strong local spectral-weight fluctuations, causing the QE to couple selectively to a smaller set of dominant bath modes.
Consequently, the QE population can exhibit enhanced oscillations, stronger non-Markovian back-action, and partial excitation trapping, all of which depend sensitively on the emitter detuning and coupling position.
In the localized regime, $V>V_{\mathrm c}$, the confinement of the bath modes limits the spread of excitation and makes the QE dynamics strongly dependent on the coupling position, leading to irregular oscillations and enhanced population retention.

We further showed that tuning the QE transition frequency enables selective coupling to distinct spectral regions of the bath.
When the QE transition frequency lies within a bath band where the LDOS is appreciable and varies smoothly, the excitation decays efficiently into the bath with no-Memory effect, whereas near band edges or spectral gaps the dynamics can develop long-lived oscillatory components with non-Markovian dynamics.
We characterized the corresponding QE--photon bound states using a self-consistent pole equation and showed that their energies and QE weights depend on the modulation strength, emitter detuning, and coupling position.
The infinite-time-averaged bound-state contribution to the QE population, obtained from the overlaps of the initial state with the bound-state eigenstates, quantifies the persistent QE excitation supported by these states.
The pronounced dependence of this contribution on $V/J$, $\Delta_{\mathrm{eg}}/J$, and the coupling site demonstrates that bound-state-mediated excitation trapping is both spectrally and spatially selective.
More importantly, these results establish the QE as a site-resolved spectroscopic probe of the quasiperiodic bath: scanning its transition frequency and coupling position reveals the global gap structure in the extended regime and the local spectral environment in the localized regime.
Beyond this spectroscopic role, quasiperiodic photonic lattices provide spatially and spectrally selective reservoirs for controlling QE dynamics.
Rather than relying solely on the bare QE--bath coupling strength, emission, non-Markovian feedback, and population trapping can be engineered through the spatial and spectral structure of the bath.
Looking forward, the coexistence of nonlocal support and strong spatial fluctuations in the multifractal critical modes may enable selective bath-mediated interactions between spatially separated emitters, opening a route to generating quantum correlations and entanglement.

\begin{acknowledgments}
This publication has emanated from research conducted with the financial support of Research Ireland, under Grants number 18/RP/6236 and 23/RC/12197 at Rinn Quantum. S. A. Z. and  Z.J.M. acknowledge TCHPC for providing computational resources. 
\end{acknowledgments}

\appendix

\setcounter{figure}{0}
\renewcommand{\thefigure}{S\arabic{figure}}

\section{\label{ap:kernel}Derivation of the QE memory-kernel equation}
In this appendix, we derive the time-nonlocal equation of motion in Eq.~\eqref{TLSdyn.eqn} by expressing the QE--AAH Hamiltonian in the bath eigenmode basis and eliminating the photonic-mode amplitudes.
Applying the transformation in Eq.~\eqref{Eq:modeTransformation}, the total Hamiltonian in the AAH eigenmode basis becomes
\begin{equation}
H =
\Delta_{\mathrm{eg}}\sigma^\dagger\sigma
+\sum_{\alpha=1}^{N}\omega_\alpha a_\alpha^\dagger a_\alpha
+\sum_{\alpha=1}^{N}
\left(
g_\alpha\sigma^\dagger a_\alpha
+g_\alpha^*\sigma a_\alpha^\dagger
\right),
\label{Eq:Heigenmode}
\end{equation}
where $g_\alpha=g\phi_{\alpha j_0} $
is the coupling amplitude between the QE and eigenmode $\alpha$.
Correspondingly, the single-excitation state in Eq.~\eqref{Eq:Wavefunction} takes the form
\begin{equation}
\ket{\psi(t)} = c_{\mathrm e}(t)\ket{e,0} +
\sum_{\alpha=1}^{N} c_\alpha(t)\ket{g,1_\alpha},
\label{Eq:WavefunctionEigenmode}
\end{equation}
where
$\ket{g,1_\alpha}:=\ket{g}\otimes a_\alpha^\dagger\ket{0}$ denotes the QE in its ground state with one photon in eigenmode $\alpha$, and
$c_\alpha(t)=\sum_{j=1}^{N}\phi_{\alpha j}^{*}c_j(t)$ is the corresponding probability amplitude.

Substituting Hamiltonian Eq.~\eqref{Eq:Heigenmode} and wave function Eq.\eqref{Eq:WavefunctionEigenmode} into the time-dependent Schr"odinger equation,
$i\partial_t\ket{\psi(t)}=H\ket{\psi(t)}$, yields the coupled amplitude equations

\begin{align}
i\dot{c}_{\mathrm e}(t)
&= \Delta_{\mathrm{eg}}c_{\mathrm e}(t)
+\sum_{\alpha=1}^{N}g_\alpha c_\alpha(t),
\\
i\dot{c}_\alpha(t)
&=
\omega_\alpha c_\alpha(t)
+g_\alpha^* c_{\mathrm e}(t).
\label{Eq:coupledAmplitudes}
\end{align}
For the initial state $\ket{\psi(0)}=\ket{e,0}$, the amplitudes satisfy $c_{\mathrm e}(0)=1$ and $c_\alpha(0)=0$. Multiplying the second equation in Eq.~\eqref{Eq:coupledAmplitudes} by the integrating factor $e^{i\omega_\alpha t}$ and integrating from $0$ to $t$ gives
\begin{equation}
c_\alpha(t) = -i g_\alpha^* \int_0^t d\tau e^{-i\omega_\alpha(t-\tau)}
c_{\mathrm e}(\tau).
\label{Eq:bathAmplitude}
\end{equation}

Substituting Eq.~\eqref{Eq:bathAmplitude} into the first equation in Eq.~\eqref{Eq:coupledAmplitudes} yields the exact integrodifferential equation
\begin{equation}
\dot{c}_{\mathrm e}(t) = -i\Delta_{\mathrm{eg}}c_{\mathrm e}(t)
-\int_0^t d\tau K_{j_0}(t-\tau)c_{\mathrm e}(\tau),
\label{Eq:eliminatedBath}
\end{equation}
Comparing Eq.~\eqref{Eq:eliminatedBath} with Eq.~\eqref{TLSdyn.eqn}, we identify the position-dependent memory kernel as
\begin{equation}
K_{j_0}(t-\tau) =
\sum_{\alpha=1}^{N}
|g_\alpha|^2e^{-i\omega_\alpha(t-\tau)} =
g^2\sum_{\alpha=1}^{N}
|\phi_{\alpha j_0}|^2e^{-i\omega_\alpha(t-\tau)}.
\label{KernelSM.eqn}
\end{equation}
Using the definitions of the LDOS in Eqs.~\eqref{Eq:LDOS} and local spectral density definition, this mode sum is equivalent to the integral representation in Eq.~\eqref{Kernel.eqn}. This completes the derivation of Eq.~\eqref{TLSdyn.eqn} without invoking a Markov approximation.

\begin{figure}[t]
\centering
\includegraphics[width=\linewidth]{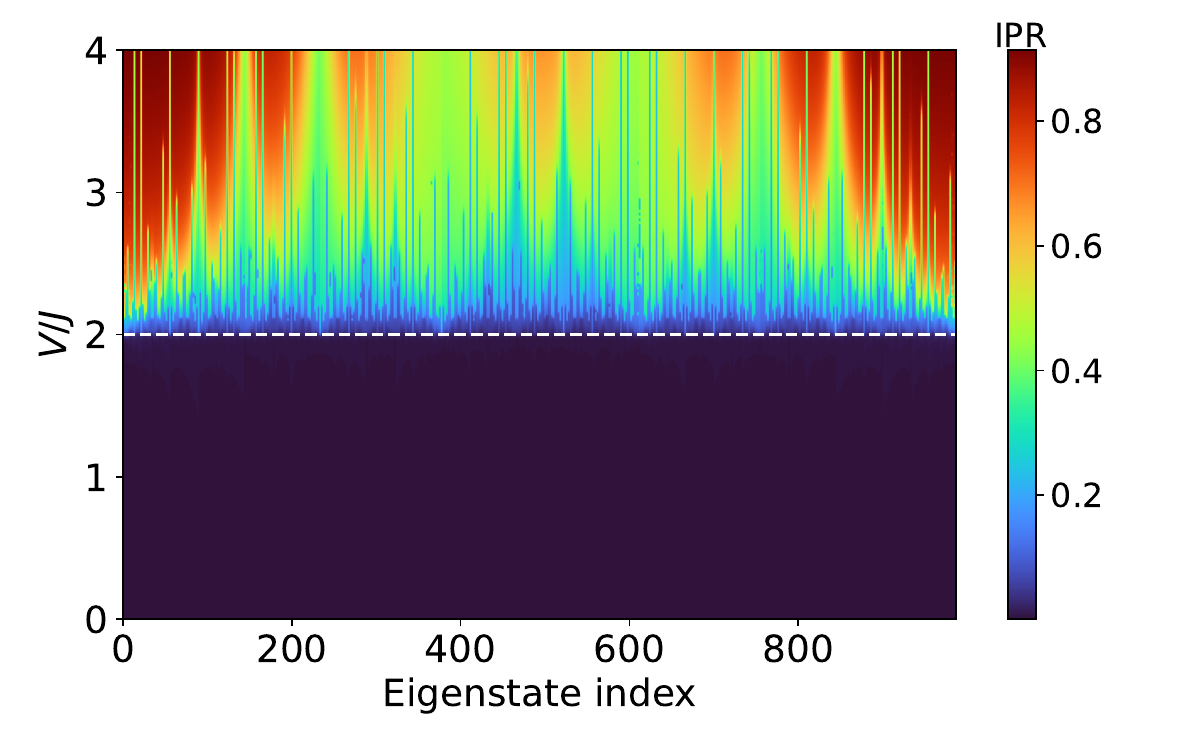}
\caption{\label{Fig2N.}
Localization properties of the AAH eigenmodes. The eigenfrequencies $\omega_\alpha/J$ are shown as a function of the normalized modulation strength $V/J$, with each eigenmode colored according to its inverse participation ratio $\mathrm{IPR}_\alpha$. Larger (smaller) IPR values correspond to more localized (extended) eigenmodes. The dashed white line marks the self-dual critical point $V_{\mathrm c}=2J$. The parameters are $N=987$, $\beta=610/987$, and $\theta=0$, with periodic boundary conditions.}
\end{figure}

\section{\label{ap:IPR}Localization properties of the AAH eigenmodes}
In this section, we verify the localization properties of the finite AAH lattice used in our calculations. The localization of each bath eigenmode $\ket{\alpha}$ is quantified through its inverse participation ratio~(IPR), defined as
\begin{equation}
\mathrm{IPR}_{\alpha} = \frac{\displaystyle\sum_{j=1}^{N}|\phi_{\alpha j}|^4}
{\left(\displaystyle\sum_{j=1}^{N}|\phi_{\alpha j}|^2\right)^2},
\label{Eq:IPR}
\end{equation}
where $\phi_{\alpha j}=\braket{1_j|\alpha}$ is the amplitude of eigenmode $\alpha$ at site $j$. Because the eigenmodes are normalized, Eq.~\eqref{Eq:IPR} reduces to
$\mathrm{IPR}_{\alpha}=\sum_{j=1}^{N}|\phi_{\alpha j}|^4$.
For an extended eigenmode distributed approximately uniformly over the entire lattice, $\mathrm{IPR}_{\alpha}\sim 1/N$ and therefore vanishes in the thermodynamic limit. By contrast, for an exponentially localized eigenmode occupying a finite number of lattice sites, the IPR remains finite as $N$ increases. At the critical point, the multifractal eigenmodes exhibit the intermediate scaling
$\mathrm{IPR}_{\alpha}\sim N^{-D_2}$, where $0<D_2<1$ is the correlation dimension.
 We approximate the incommensurability parameter in the finite lattice by the ratio of consecutive Fibonacci numbers, $\beta=610/987$. We consequently take $N=987$ and set the global phase to $\theta=0$. For this bulk localization diagnostic, periodic boundary conditions are imposed, with $b_{N+1}=b_1$. The rational approximation $\beta$ then makes the quasiperiodic onsite potential compatible with the finite periodic lattice.
Figure~\ref{Fig2N.} reproduces the established extended-to-localized transition of the standard AAH model and provides a finite-size consistency check for the three regimes examined in the main text.

\begin{figure}[t]
    \centering
    \includegraphics[width=1 \linewidth]{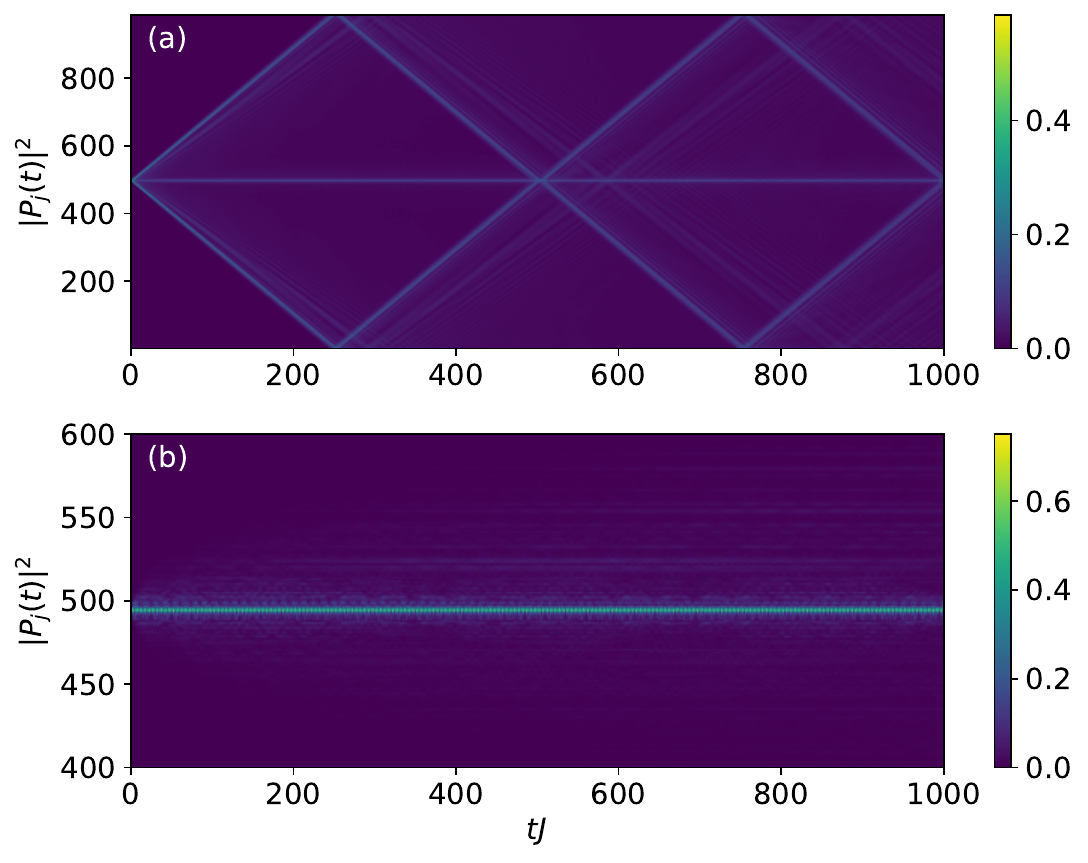}
\caption{Spatiotemporal evolution of the site-resolved photonic population $P_j(t)=|c_j(t)|^2$ following excitation of a QE coupled to the central site $j_0=[N/2]$ of an AAH lattice with periodic boundary conditions. The QE is resonant with the bare cavity frequency,$\Delta_{\mathrm{eg}}=0$. (a) In the extended regime, $V=0.1J$, counterpropagating wave fronts traverse the periodic lattice and subsequently return to the coupling site, producing a finite-size recurrence. (b) At the critical point, $V=V_{\mathrm c}=2J$, the photonic population remains concentrated near $j_0$ over the displayed time interval, reflecting strongly inhibited propagation associated with the multifractal bath modes. The color scale represents $P_j(t)$.
}
\label{Fig.S2p}
\end{figure}

\section{Excitation propagation and finite-size recurrences}
\label{ap:recurrence}

In this section, we examine the propagation of the photonic excitation through the AAH lattice and determine how the finite-size recurrence time depends on the lattice size in the extended regime.
We initialize the system in the state $\ket{\psi(0)}=\ket{e,0}$, couple the QE to the central site $j_0 = [N/2]$ of an AAH chain with periodic boundary conditions, and monitor the site-resolved photonic population $P_j(t)=|c_j(t)|^2$.
In the extended regime, the emitted excitation forms counterpropagating wave fronts that spread ballistically away from $j_0$, as evidenced by the approximately linear light-cone boundaries in Fig.~\ref{Fig.S2p}(a).
As a result of periodic boundary conditions, these wave fronts propagate around the lattice and subsequently return to the coupling site at $t=T_{\mathrm{rec}}$, producing the finite-size recurrence observed in the QE dynamics.
At the critical point, by contrast, the excitation remains concentrated within an inhomogeneous region around coupling position $j_0$ over the displayed time interval, reflecting the strongly inhibited spreading associated with multifractal bath modes.
This persistent photonic weight near the coupling region maintains substantial overlap with the QE and thereby produces the early bath back-action, non-Markovian dynamics and population revivals observed in Fig.~\ref{Fig3} and~\ref{Fig4}.
Thus, unlike the finite-size recurrence in the extended regime, the critical-regime revivals originate from local coupling to multifractal bath modes rather than from an excitation traversing the entire lattice.

\begin{figure}[t]
    \centering
    \includegraphics[width=1 \linewidth]{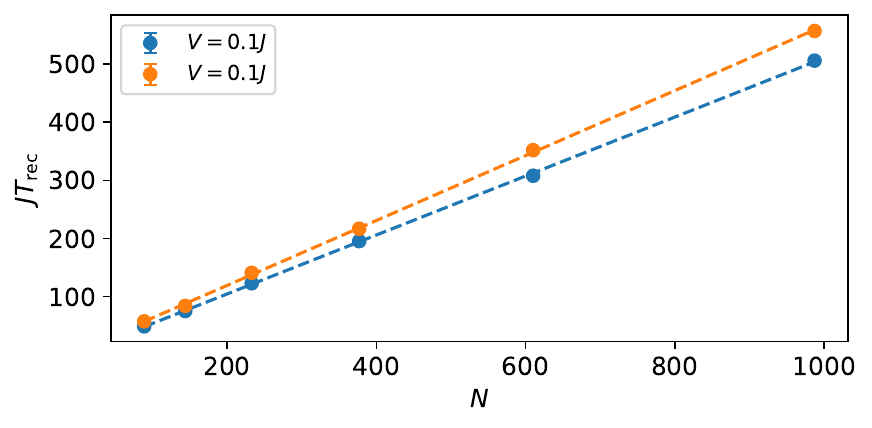}
    \caption{\label{Fig:RecurrenceScaling}
Finite-size scaling of the QE recurrence time in the extended regime. The recurrence time $T_{\mathrm{rec}}$ is identified as the first pronounced local maximum of $P_{\mathrm e}(t)$ following its initial decay and is shown as a function of the lattice size $N$ for $V=0.1J$ and $V=0.5J$. The symbols denote the numerical results, and the dashed lines are linear least-squares fits. The vertical error bars represent the temporal resolution of each calculation, $\delta T_{\mathrm{rec}}=d t_N/2$. The approximately linear dependence $T_{\mathrm{rec}}\propto N$ confirms that the revival originates from ballistic propagation around the finite periodic lattice. The QE is coupled to the central site, and $\Delta_{\mathrm{eg}}=0$.}
\label{FigS3p}
\end{figure}

To verify the finite-size origin of the recurrence, we next calculate the QE dynamics for several lattice sizes at fixed modulation strength $V=0.1J$.
We define $T_{\mathrm{rec}}$ as the time of the first pronounced revival of $P_{\mathrm e}(t)$ following its initial decay, corresponding to the return of the propagating excitation to the coupling site.
Figure~\ref{FigS3p} shows the extracted recurrence time for the Fibonacci lattice sizes $N=89$, $144$, $233$, $377$, $610$, and $987$ at $V=0.1J$ and $V=0.5J$, both within the extended regime. For each $N=F_m$, the incommensurability parameter is approximated by the ratio of consecutive Fibonacci numbers, $\beta=F_{m-1}/F_m$.
Because $T_{\mathrm{rec}}$ is extracted from a discrete time grid with a size-dependent spacing $d t_N$, the vertical error bars represent the corresponding temporal resolution, $\delta T_{\mathrm{rec}}=d t_N/2$.
For both modulation strengths, $T_{\mathrm{rec}}$ increases approximately linearly with $N$, confirming that the revival is a finite-size propagation effect rather than a local reservoir-memory process.
Unweighted least-squares fits yield $JT_{\mathrm{rec}}\simeq 3.06+ 0.507N$ for $V=0.1J$ and $JT_{\mathrm{rec}}\simeq 7.53 + 0.559N$ for $V=0.5J$.
The approximately linear scaling $T_{\mathrm{rec}}\propto N$ confirms that the recurrence originates from the ballistic traversal of the finite periodic lattice. The fitted slopes correspond to effective propagation velocities $v_{\mathrm{eff}}\simeq1.97J$ for $V=0.1J$ and $v_{\mathrm{eff}}\simeq1.79J$ for $V=0.5J$, taking the lattice spacing as unity.
Consequently, $T_{\mathrm{rec}}$ diverges in the thermodynamic limit, distinguishing these finite-size recurrences from the short-time, position-dependent bath feedback observed at the critical point.

\section{Decay analysis}
\label{Decay}

For amplitude-damping dynamics, we define the QE excited-state
population as $p_{\mathrm e}(t)=|c_{\mathrm e}(t)|^2$. Whenever
$p_{\mathrm e}(t)>0$, the corresponding time-local decay rate is
\begin{equation}
    \gamma(t)
    =
    -\frac{\dot{p}_{\mathrm e}(t)}{p_{\mathrm e}(t)}.
    \label{Eq:instantaneous_decay_rate}
\end{equation}
Accordingly, $\gamma(t)>0$ describes excitation loss from the QE,
whereas $\gamma(t)<0$, equivalently $\dot{p}_{\mathrm e}(t)>0$,
identifies intervals of population revival associated with excitation
backflow. Care must be taken near zeros of $p_{\mathrm e}(t)$, where
Eq.~\eqref{Eq:instantaneous_decay_rate} becomes singular and can
produce large numerical spikes.

To quantify the accumulated population revival over the observation
interval $[0,T]$, we introduce the nonnegative quantities~\cite{PhysRevA.111.022212}
\begin{align}
    \mathcal{N}_{+}(T)
    &=
\int_{\substack{ \dot{p}_{\mathrm e}(t)>0}}
    dt\,
    \frac{d p_{\mathrm e}^{2}(t)}{dt},
    \\
    \mathcal{N}_{-}(T)
    &=
    -\int_{\substack{ \dot{p}_{\mathrm e}(t)<0}}
    dt\,
    \frac{d p_{\mathrm e}^{2}(t)}{dt},
    \label{Eq:positive_negative_variations}
\end{align}
where
\begin{equation}
    \frac{d p_{\mathrm e}^{2}(t)}{dt}
    =
    2p_{\mathrm e}(t)\dot{p}_{\mathrm e}(t).
\end{equation}
Here, $\mathcal{N}_{+}$ measures the total increase of
$p_{\mathrm e}^{2}(t)$ during population-revival intervals, while
$\mathcal{N}_{-}$ measures its total decrease. Their ratio,
\begin{equation}
    \chi(T)
    =
    \frac{\mathcal{N}_{+}(T)}{\mathcal{N}_{-}(T)},
    \label{Eq:nonMarkovianity_ratio}
\end{equation}
provides a normalized measure of population backflow. For an initially
excited QE, $0\leq\chi(T)\leq1$, provided
$\mathcal{N}_{-}(T)>0$. A value $\chi(T)=0$ indicates monotonic decay
and hence the absence of population-revival signatures of
non-Markovianity, whereas larger values indicate stronger revivals.
The limit $\chi(T)\rightarrow1$ corresponds to nearly recurrent
dynamics in which the accumulated population recovery approaches the
accumulated population loss.

\begin{figure}[t]
    \centering
    \includegraphics[width=1
    \linewidth]{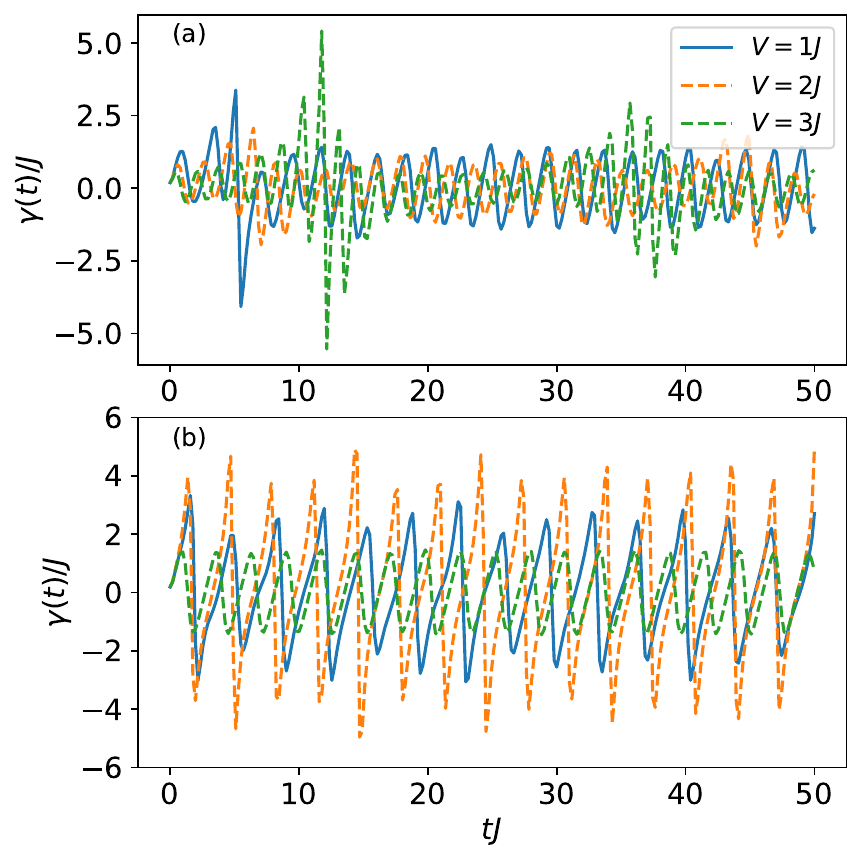}
    \caption{QE dynamics loss under different on-site potentials and transition frequencies. In panels~(a) and~(b), $\omega_\mathrm{eg} =0,2$ is assumed.  The lattice size is $N=987$ and the position of emitter is $N$. The last row indicates the phase portrait of the results in panels~(c), and~(d) in the same color code as in panel~(a). }
    \label{FigS4}
\end{figure}

\bibliography{aipsamp}

\end{document}